\documentclass[a4paper]{article}
\usepackage{bm}
\usepackage{amsmath,amssymb}
\usepackage{url}
\usepackage{caption}
\usepackage[dvipdfmx]{graphicx}
\usepackage[dvipdfmx]{color}
\usepackage{arydshln}
\usepackage{mathtools}
\usepackage[top=2.85cm, bottom=3.25cm, left=2.45cm, right=2.45cm]{geometry}
\usepackage[colorlinks=true, linkcolor=blue, urlcolor=blue]{hyperref}
\usepackage{colortbl}
\usepackage[table,xcdraw]{xcolor}
\usepackage{xcolor}

\begin{document}
\baselineskip=1.45\baselineskip

\begin{center}
{\Large
\textbf{Optimal Bayesian predictive probability for delayed response}
}\\\vspace{4pt}
{\Large
\textbf{in single-arm clinical trials with binary efficacy outcome}
}
\end{center}

\

\begin{center}
Takuya Yoshimoto$^{1, 2}$, Satoru Shinoda$^{2}$, Kouji Yamamoto$^{2}$ and Kouji Tahata$^{3}$

\

$^{1}${Biometrics Department, Chugai Pharmaceutical Co., Ltd., Chuo-ku, Tokyo, 103-8324, Japan}\\
$^{2}${Department of Biostatistics, Yokohama City University School of Medicine, Yokohama City, Kanagawa, 236-0004, Japan}\\
$^{3}${Department of Information Sciences, Faculty of Science and Technology, Tokyo University of Science, Noda City, Chiba, 278-8510, Japan}

\

E-mail: \href{mailto:yoshimoto.takuya61@chugai-pharm.co.jp}{\nolinkurl{yoshimoto.takuya61@chugai-pharm.co.jp}}

\

Last update: May 18, 2024
\end{center}
\

\begin{abstract}
In oncology, phase II or multiple expansion cohort trials are crucial for clinical development plans. This is because they aid in  identifying potent agents with sufficient activity to continue development and confirm the proof of concept. Typically, these clinical trials are single-arm trials, with the primary endpoint being short-term treatment efficacy. 
Despite the development of several well-designed methodologies, there may be a practical impediment in that the endpoints may be observed within a sufficient time such that adaptive go/no-go decisions can be made in a timely manner at each interim monitoring.
Specifically, Response Evaluation Criteria in Solid Tumors guideline defines a confirmed response and necessitates it in non-randomized trials, where the response is the primary endpoint. However, obtaining the confirmed outcome from all participants entered at interim monitoring may be time-consuming as non-responders should be followed up until the disease progresses.
Thus, this study proposed an approach to accelerate the decision-making process that incorporated the outcome without confirmation by discounting its contribution to the decision-making framework using the generalized Bayes' theorem. Further, the behavior of the proposed approach was evaluated through a simple simulation study. 
The results demonstrated that the proposed approach made appropriate interim go/no-go decisions.

\end{abstract}

\noindent
$\textbf{Keywords}$: Bayesian monitoring, binary outcome, delayed response, predictive probability, quasi-posterior distribution, single-arm clinical trials.

\hypertarget{S1}{}
\section{Introduction}
In oncology, the phase II trials or multiple expansion cohort trials (FDA \hyperlink{R27}{Guidance for industry, 2022}), which are intended to expedite development by seamlessly proceeding from determination of a tolerated dose to assessments that are more typical of phase II trials, are aimed at the evaluation of the preliminary efficacy of a new treatment to identify agents with sufficient activity to confirm the proof of concept (PoC). Typically, these types of clinical trials follow a single-arm trial, with the primary endpoint being short-term treatment efficacy. Specifically, the objective response based on the Response Evaluation Criteria in Solid Tumors (RECIST) (\hyperlink{R6}{Eisenhauer et al., 2009}) guideline is commonly used as a primary endpoint in terms of the treatment efficacy. These trials routinely implement futility monitoring and interim go/no-go decision-making to protect participants against ineffective treatments and to accelerate the clinical development. Several frequentist and Bayesian designs have been proposed for single-arm clinical trials with binary outcomes. 
One such well-known frequentist framework design is the Simon's two-stage design (\hyperlink{R23}{Simon, 1989}) that minimizes the maximum sample size or the expected sample size under the null hypothesis that the treatment is ineffective, while also controlling for type I and II error rates. \hyperlink{R7}{Fleming (1982)} suggested that trials should be stopped early in case the results of the first stage are extreme, either in favor of efficacy or futility. Compared with Simon's design, early acceptance of the experimental treatment is permitted. For the Bayesian framework, \hyperlink{R25}{Thall and Simon (1994)} provided practical guidelines for the implementation of adaptive single-arm clinical trials that involved updating the Bayesian posterior probability after observing the data. \hyperlink{R11}{Lee and Liu (2008)} developed a predictive probability-based Bayesian monitoring scheme. They evaluated the probability of achieving the desired outcome at the planned end of the trial based on accumulated interim data. 
In addition, \hyperlink{R21}{Sambucini (2021)} extended the designs of \hyperlink{R11}{Lee and Liu (2008)} by considering the uncertainty of the historical control.
The predictive distribution approach considers decisions upon final analysis based on data to be available in the future, whereas the posterior distribution approach is reliant only on data at intermediate time points. \hyperlink{R22}{Saville et al. (2014)} mentioned that definitive conclusions are often difficult to obtain from data at interim time points; therefore, the effectiveness of the final analysis demands that the data that will be available in the future should be considered. Therefore, predictive distributions are easy to interpret as decision-making at the time of the final analysis. In practice, several clinical trials have been designed using the Bayesian approach (see, e.g., \hyperlink{R10}{Hirakawa et al. 2018}; \hyperlink{R5}{Domchek et al., 2020}; \hyperlink{R14}{Murai et al., 2021)}.

A major practical impediment to the implementation of a continuous monitoring is that the endpoints may be observed within a sufficient time such that adaptive go/no-go decisions can be made in a timely manner at each interim monitoring.
Specifically, RECIST guideline defines confirmed response as complete response (CR) or partial response (PR) on two consecutive tumor assessments at least 4 weeks apart. Further, it states that in non-randomized trials (where the response is the primary endpoint), confirmed CR or PR is needed to deem either one.  
Hereafter, the outcome with confirmation and the outcome without non-confirmation are referred to as the \textit{confirmed outcome} and \textit{non-confirmed outcome} throughout this paper. 
In practice, obtaining the confirmed outcome from all participants entered at interim monitoring may be time-consuming as non-responders should be followed up until the disease progresses.
One intuitive idea to address this issue requires that decision-making should be postponed at the interim to wait for the data to be confirmed. 
However, this approach may be infeasible in practice because repeated suspension of participant accruals may result in unacceptably long trials, as mentioned above. Another pragmatical approach involves decision-making performed after all participants enrolled at the timing of interim monitoring have completed at least two tumor assessments.
Although this approach can accelerate clinical development, participants who did not achieve a confirmed response at the time of analysis were considered non-responders even if they achieved CR or PR as the non-confirmed outcome. Therefore, the potential of the experimental treatment may have been  underestimated.  
To address this issue, it is not surprising to consider the non-confirmed outcome should be incorporated into a decision-making framework, as anti-cancer agents work by decreasing the tumor burden via the shrinking of the tumor. 
In the context of phase I clinical trials, certain designs incorporate information at the interim monitoring to likelihood by weighting the probability of response with a proportion of the time to interim monitoring to the assessment period (e.g., \hyperlink{R4}{Cheung and Chappell, 2000}; \hyperlink{R24}{Takeda et al., 2020}) considering the time to response outcome as being uniformly distributed over the specific assessment period. Although these designs are beneficial, they are not applicable to the situations wherein the assessment period cannot be reasonably pre-specified, as is the case with RECIST. \hyperlink{R1}{Asakawa and Hamada (2013)} highlighted this issue and proposed an approach that uses early efficacy responses without confirmation as the surrogate outcome for the confirmed outcome. 
Determining the next dose is a barrier to participant enrollment in phase I clinical trials; correspondingly, this process exerts a high impact on the duration of the entire study. On the other hand, in phase II or phase Ib trials, such as multiple expansion cohort trials, the decision-making does not become a bottleneck in terms of participant enrollment, as the single-arm trials with a fixed dose are often designed. However, an interim monitoring remains vital from the perspective of accelerating clinical development, and eliminating the influences of a delayed outcome to the best extent possible.
Accordingly, based on an idea proposed by \hyperlink{R1}{Asakawa and Hamada (2013)}, we focus on the improvement of the design of single-arm clinical trials concerning the study duration to avoid performance degradation compared with those of the foregoing approaches.
Specifically, we consider the use of the non-confirmed outcome as the surrogate outcome for the confirmed outcome, because a moderate positive correlation is expected. Subsequently, we then focus on an approach that incorporates the non-confirmed outcome by discounting its contribution to the decision-making framework using the generalized Bayes' theorem.
Thus, our primary objectives are to (i) propose a new approach for incorporating the non-confirmed outcomes into the decision-making based on a predictive distribution and (ii) evaluate the behavior of the proposed approach using a simple simulation study. 

The remainder of this paper is organized as follows. In Section \hyperlink{S2}{2}, we introduce the notation and propose a new approach to incorporate the non-confirmed outcome into the decision-making process. In Section \hyperlink{S3}{3}, we present the results of a simulation study aimed at evaluating the operating characteristics of the proposed methodology in conjunction with the design parameter optimization. 
Finally, Section \hyperlink{S4}{4} presents the discussion.
All the computations presented in this study were performed using R programming (\hyperlink{R20}{R Core Team 2023}).

\hypertarget{S2}{}
\section{Methods}
\hypertarget{S21}{}
\subsection{Approach considering the delayed response}
Herein, we represent the efficacy of an experimental treatment, E, and assume that a standard treatment, S, is collected from historical control data obtained as an aggregated data. The maximum sample size planned for the entire trial of E is denoted by $N_{max}$. The observed Bernoulli outcome of participants who have entered the trial for E is denoted by $x_{i}$ for the $i$th participant, which indicates whether the participant has achieved a response ($x_{i}=1$) or not ($x_{i}=0$) by the time of decision-making.
Let $\gamma_{i}$ be denoted by the response outcome that has been ascertained as confirmation ($\gamma_{i}=1$) or is still pending ($\gamma_{i}=0$) by the decision time. Then, $i$th-participant with $x_{i}=1$ and $\gamma_{i}=1$ indicates that this participant obtained a confirmed response at the decision time.
Further, $n$ ($n \leq N_{max}$) denotes the total number of participants at a certain interim time. Herein, $t$ = E and S denote the index treatment.
Subsequently, we introduce a beta prior for the response rate $p_{t}$,
$$
p_{t}|\alpha_{t}, \beta_{t} \sim beta(\alpha_{t}, \beta_{t}),
$$
where $\alpha_{t}$ and $\beta_{t}$ are the hyperparameters of $p_{t}$. We then consider the information from the interim monitoring for E. To consider the non-confirmed outcome in the decision-making, it is natural to discount its contribution to the information, say $w$ $(0<w<1)$. This is because a response from the non-confirmed outcome may be inconsistent with a response from the confirmed outcome.
Accordingly, these non-confirmed outcomes can be considered as fractional events. 
Therefore, through the addition of the manipulation of discounting, no binary outcome is available for a responder or non-responder because the non-confirmed outcome represents the values of the real number $w \in (0,1)$. In this case, the probability distribution is no longer binomial. Thus, the likelihood cannot be obtained to update the posterior distribution using Bayes' theorem. Instead, we consider the following generalized Bayes' theorem:
\hypertarget{E1}{}
\begin{equation}
\displaystyle f \left(p_{E}|\alpha_{E}, \beta_{E}, X\right)=\frac{{\rm exp}\left\{-l(p_{E}, X)\right\}f \left(p_{E}|\alpha_{E}, \beta_{E}\right)}{\displaystyle\int_{0}^{1}{\rm exp}\left\{-l(p_{E}, X)\right\}f \left(p_{E}|\alpha_{E}, \beta_{E}\right)dp_{E}},
\end{equation}
where $X$ is the response information obtained at the interim monitoring, that is, $X=(x_{i}, \gamma_{i})$ for all $i=1,\dots, n$, $l(p_{E}, X)$ is a loss function, $f \left(p_{E}|\alpha_{E}, \beta_{E}\right)$ is a prior distribution of $p_{E}$, and $f \left(p_{E}|\alpha_{E}, \beta_{E}, X\right)$ is a quasi-posterior distribution of $p_{E}$. Although \hyperlink{R2}{Bissiri et al. (2016)} characterized the general Bayesian approach using a set of assumptions or axioms, the result was restricted to a finite probability space. Thereafter, \hyperlink{R3}{Bissiri and Walker (2019)} extended these results to a real line by considering a Borel measurable subset. Specifically, to address the uncountable case, \hyperlink{R3}{Bissiri and Walker (2019)} considered another request of the updated probability not being affected by the removal of a set with zero prior probability from the parametric space, in addition to the axiom described in \hyperlink{R2}{Bissiri et al. (2016)}.
In addition, they demonstrated that these axioms uniquely implied a generalized Bayes equation. Notably, Equation (\hyperlink{E1}{1}) reduces the well-known Bayes' theorem when the likelihood function is set to the loss function. Using the generalized Bayes' theorem, we can consider the quasi-posterior distribution of $p_{E}$.
First, we now consider the joint probability at the time of interim monitoring:
$$
\displaystyle\prod_{i=1}^{n}p_{E}^{\gamma_{i}x_{i}+w(1-\gamma_{i})x_{i}}\left(1-p_{E}\right)^{\gamma_{i}(1-x_{i})+w(1-\gamma_{i})(1-x_{i})}.
$$
From the joint probability, we set the loss function $l(p_{E}, X)$ as follows:
$$
l(p_{E}, X) =-\left(\sum_{i=1}^{n}\gamma_{i}x_{i} + w(1-\gamma_{i})x_{i}\right)\log p_{E} - \left(\sum_{i=1}^{n}\gamma_{i}(1-x_{i})+w(1-\gamma_{i})(1-x_{i})\right)\log (1 - p_{E}).
$$
Consequently, we obtain the following quasi-posterior distribution from Equation (\hyperlink{E1}{1}):
$$
\displaystyle p_{E}|\alpha_{E}, \beta_{E}, X \sim beta\left(\alpha_{E}^{*}, \beta_{E}^{*}\right),
$$
where $\alpha_{E}^{*}=\alpha_{E} + \sum_{i=1}^{n}\gamma_{i}x_{i} + w(1-\gamma_{i})x_{i}$ and $\beta_{E}^{*}=\beta_{E} + \sum_{i=1}^{n}\gamma_{i}(1-x_{i})+w(1-\gamma_{i})(1-x_{i})$. 
We can easily calculate the dominator of Equation (\hyperlink{E1}{1}) using the normalized constant of $beta(\alpha_{E}^{*}, \beta_{E}^{*})$.

Subsequently, we consider the posterior probability that $p_{E}$ exceeds the $p_{S}$, based on \hyperlink{R25}{Thall and Simon (1994)} and \hyperlink{R21}{Sambucini (2021)}. 
The experimental treatment is considered sufficiently promising if ${\rm Pr} \left(p_{E} > p_{S}|\alpha_{E}^{*}, \beta_{E}^{*}, \alpha_{S}, \beta_{S}\right)$. Therefore, the posterior probability that the experimental treatment is worthy can be computed as
\hypertarget{E2}{}
\begin{equation}
{\rm Pr} \left(p_{E} > p_{S}|\alpha_{E}^{*}, \beta_{E}^{*}, \alpha_{S}, \beta_{S}\right)=\int_{0}^{1}\left[1- Beta(p_{S}; \alpha_{E}^{*}, \beta_{E}^{*})\right]beta(p_{S}; \alpha_{S}, \beta_{S})dp_{S},
\end{equation}
where $beta(\cdot; \alpha_{E}^{*}, \beta_{E}^{*})$ is the probability density function and $Beta(\cdot; \alpha_{E}^{*}, \beta_{E}^{*})$ is the cumulative distribution function of the beta distribution with parameters $\alpha_{E}^{*}$ and $\beta_{E}^{*}$, respectively (see Appendix \hyperlink{A1}{A} for additional details). The use of a prior distribution for $p_{S}$ allows the incorporation of uncertainty into the historical response rate of the standard treatment.

\hypertarget{S22}{}
\subsection{Predictive monitoring}
We present the decision-making process based on the predictive probability. Let $Y$ be a random variable representing the number of confirmed responses out of the potential future $m$, that is, $m=N_{max}-n$, participants. When the result $Y = y$ is available, we construct the following quasi-posterior distribution for $p_{E}$: 
$$
p_{E}|\alpha_{E}^{*},\beta_{E}^{*},y \sim beta(\alpha_{E}^{*}+y,\beta_{E}^{*}+m-y).
$$
In addition, the posterior predictive distribution of $Y$ is
\hypertarget{E3}{}
\begin{equation}
f_{m}\left(y|\alpha_{E}^{*}, \beta_{E}^{*}\right)=\displaystyle {m \choose y}\frac{B(\alpha_{E}^{*}+y, \beta_{E}^{*}+m-y)}{B(\alpha_{E}^{*}, \beta_{E}^{*})},
\end{equation}
where $B(\cdot,\cdot)$ is the beta function and ${m \choose y}$ is the binomial coefficient. This is a beta compound binomial distribution. At the conclusion of the trial, the experimental treatment is declared sufficiently promising if the following condition based on Equation (\hyperlink{E2}{2}) is satisfied:
$$
{\rm Pr} \left(p_{E} > p_{S}|\alpha_{E}^{*}, \beta_{E}^{*}, \alpha_{S}, \beta_{S}, y\right) > \lambda,
$$
where $\lambda$ is a pre-specified probability threshold, and
\hypertarget{E4}{}
\begin{equation}
{\rm Pr} \left(p_{E} > p_{S}|\alpha_{E}^{*}, \beta_{E}^{*}, \alpha_{S}, \beta_{S}, y\right)=\int_{0}^{1}\left[1- Beta(p_{S}; \alpha_{E}^{*}+y, \beta_{E}^{*}+m-y)\right]beta(p_{S}; \alpha_{S}, \beta_{S})dp_{S}.
\end{equation}
Subsequently, $I(\cdot)$ is considered as an indicator that maps the elements of the subset to one, and all other elements to zero. From Equations (\hyperlink{E3}{3}) and (\hyperlink{E4}{4}), we can compute the predictive probability that the experimental treatment is declared sufficiently promising at the scheduled end of the trial using the following equation:
\hypertarget{E5}{}
\begin{equation}
PP = \sum_{y}f_{m}\left(y|\alpha_{E}^{*}, \beta_{E}^{*}\right)I\left({\rm Pr} \left(p_{E} > p_{S}|\alpha_{E}^{*}, \beta_{E}^{*}, \alpha_{S}, \beta_{S}, y\right) > \lambda\right).
\end{equation}
This is interpreted as the weighted average of the indicators of a successful trial upon the end of the trial. It is based on the probability that the experimental treatment is considered sufficiently promising. A high $PP$ indicates that the experimental treatment is likely to be efficacious by the end of the trial considering the current data. In contrast, a low $PP$ suggests that the experimental treatment may not offer sufficient activity. In line with \hyperlink{R11}{Lee and Liu (2008)} and \hyperlink{R21}{Sumbucini (2021)}, the decision rules are constructed as follows:
\begin{itemize}
\setlength{\parskip}{0cm} 
\setlength{\itemsep}{0cm}
\item if $PP < \theta_{L}$, then stop the trial and declare that the experimental treatment is not sufficiently successful;
\item if $PP > \theta_{U}$, then stop the trial and declare that the experimental treatment is sufficiently successful;
\item otherwise, continue enrolling participants until the maximum sample size is reached.
\end{itemize}
Typically, $\theta_{L}$ is selected as a small positive constant and $\theta_{U}$ as a large positive constant, with values between $0$ and $1$. $PP < \theta_{L}$ indicates that the trial is unlikely to yield the desired performance in terms of efficacy at the end of the trial given the current information. However, when $PP > \theta_{U}$, the current data suggest a high probability of concluding that the experimental treatment is efficacious at the end of the trial.

\hypertarget{S3}{}
\section{Operating characteristics of the proposed design and threshold calibration}
\hypertarget{S31}{}
\subsection{Simulation setting}
In this section, we conduct simulations to investigate the performance of the proposed Bayesian interim procedure empirically in terms of the frequentist operating characteristics. 
Although futility analyses are commonly planned in single-arm clinical trials with the primary endpoint being treatment efficacy in terms of ethics (\hyperlink{R9}{Green et al., 2012}), the clinical development of more effective treatments must be accelerated. This is consistent with the concept of incorporating the non-confirmed outcome into the decision-making.  In rare diseases, the time required to complete the enrollment of a fixed number of participants is prolonged in certain cases owing to their rarity. For example, \hyperlink{R10}{Hirakawa et al. (2018)} applied a Bayesian design with pre-specified boundaries only considering efficacy reasons. Therefore, to evaluate the PoC of our new approach, we set $\theta_{L}=0$ such that the trial can be terminated only because of its high efficacy, and set $\theta_{U}$ = 0.8. In addition, we set $\alpha_{E}=\beta_{E}=0.5$ as Jeffrey's prior with a vague prior. The beta prior density for $p_{S}$ was informatively assumed to be $beta(20, 80)$, with a mean of $0.2$ and an effective sample size of $100$.
We assumed that the first interim analysis was conducted after observing $N_{min}=15$ participants; subsequently, the data were monitored using a cohort size of five until the maximum sample size $N_{max}=40$ is reached. In addition, the null ($H_{0}$) and alternative ($H_{1}$) hypotheses were set to evaluate the behavior of the proposed approach in various scenarios:
$$
H_{0}:~{p}_{E}={p}_{S}\quad {\rm vs.}\quad H_{1}:~{p}_{E}\not={p}_{S}.
$$
To evaluate the design performance, we considered the following four metrics: 
(i) The proportion of early terminations (PET) is defined as the proportion of trials that are terminated early.
(ii) The proportion of rejection of the null hypothesis (PRN) is defined as the proportion of the simulated trials wherein $H_{0}$ is rejected, that is, the frequency of simulated trials that are stopped regardless whether the maximum sample size is reached. The PRN is the type I error rate (or power) when $H_{0}$ is (or is not) true, or the percentage of claims that the experimental treatment is effective.
(iii) The actual sample size is defined as the average sample size (ASS) used in 10000 simulated trials.
(iv) The actual study duration is defined as the average study duration (ASD) of 10000 simulated trials.

Herein, the response rate based on the confirmed outcome and the response rate that do not require a confirmatory assessments are referred to as the \textit{best objective response rate} and \textit{best response rate}, respectively.
We derived the correlated best response and objective best response rates by using the bivariate Bernoulli distribution (further details presented in Appendix \hyperlink{A2}{B}). In our simulations, the correlation coefficient between these outcomes was assumed to be $0.5$. The design parameters of $\lambda$, $\theta_{U}$ and $w$ were set to $0.8$, $0.8$ and $0.5$, respectively. The participants were uniformly enrolled with ramp-up recruitment, that is, the first $15$ ($N_{min}$) participants were recruited within $15$ months and afterwards. Thereafter, $5$ participants (per cohort) were recruited within $3$ months.
For generating the correlated bivariate Bernoulli distribution, we used \texttt{bindata} package (\hyperlink{R12}{Leisch et al., 1998}). In addition, the \texttt{ph2bayes} package (\hyperlink{R15}{Nagashima, 2018}) was used to calculate Equation (\hyperlink{E4}{4}). Furthermore, surrogate efficacy outcome with non-confirmation was assumed to be uniformly observed within the fourth tumor assessments and confirmation was derived based on the assumption that the time from obtaining the surrogate outcome to confirmation should be normally distributed such that the confirmation was obtained at the first (subsequent) tumor assessment with a probability of approximately $68\%$ and the second (subsequent) tumor assessments with a probability of approximately $95\%$. Tumor assessments should be performed every $8$ weeks.
To examine the utility of the proposed approach, we compared the two approaches mentioned in Section \hyperlink{S1}{1}. Specifically, the method is that a decision-making should be postponed at the interim monitoring to wait for the data to be confirmed. Another method involves the execution of the decision-making process after all participants enrolled at the time of interim monitoring had completed at least two tumor assessments. Hereafter, these approaches are referred to as the $\textit{performance-oriented approach}$ and $\textit{expedition-oriented approach}$, respectively. The performance- and expedition-oriented approaches are based on the method proposed by \hyperlink{R21}{Sambucini (2021)}. Appendix \hyperlink{A3}{C} presents further details on the simulation algorithm of the proposed approach. 

\hypertarget{S32}{}
\subsection{Simulation results}
Compared with the performance-oriented approach, we expected that the proposed approach could provide similar results in terms of PET, PRN and ASS; however, the proposed approach could shorten ASD. In addition, we expected that the proposed approach could provide promising results in terms of the all four performance metrics compared with the expedition-oriented approach. Table \hyperlink{T1}{1} lists the operating characteristics of these three approaches, summarized by the empirical probabilities of PET, PRN, ASS and ASD. 

\begin{center}
\lbrack Table 1 about here.\rbrack
\end{center}

\noindent
Consider the case with Scenarios 1, 2 and 3 underlying the $H_{1}$ corresponding to the cases with small and large differences between the best objective response and the best response rates. Regardless of the true value of the best objective response rate based on the confirmed outcome, the proposed approach resulted in a loss of power compared with the performance-oriented approach; however, the magnitude was considerably small and substantially shortened the ASD.
In Scenario 1, wherein the true best objective response rate was $0.4$, the expedition-oriented approach was considerably inferior to the performance approach in terms of statistical power. However, the ASD was considerably shorter as expected. When the true best objective response rate was $0.6$ as in Scenario 3, the PRN of the expedition-oriented approach was substantially lower than that of the performance-oriented approach. Whereas, that of the performance-oriented approach was relatively high. This results in a situation wherein the ASD of the expedition-oriented approach tended to be longer among the three approaches.
The same considerations were applicable to the PET. 
In terms of the ASS, a lower PET indicates a higher frequency of interim monitoring; therefore, the expedition-oriented approach tended to have a relatively higher ASS.
Furthermore, considering Scenarios 4, 5 and 6 underlying the $H_{0}$, Scenarios 4 and 5 also corresponded to the cases wherein there were minor differences between the best objective response and the best response rates. Further, Scenario 6 corresponded to the case wherein there was no difference between these response rates. 
Based on the $H_{0}$ hypothesis, the extent of type I error inflation must be lowered compared with that of the performance-oriented approach. 
The proposed approach tended to facilitate minor increases of the type I error compared with the performance-oriented and expedition-oriented approaches when the best response rate was equal to $0.3$ or $0.4$, as in Scenarios 4 and 5. 
However, the inflation of type I errors exerted a relatively minor impact on this simulation study; thus, all approaches could efficiently control the overall type I error.

\subsection{Threshold calibration}
In Sections \hyperlink{S31}{3.1} and \hyperlink{S32}{3.2}, the tuning parameters were set to a certain value to evaluate the PoC of the proposed method and to avoid comparing methods in favorable situations with the proposed approach. However, the optimization of these tuning parameters is important for practical use. Therefore, we also illustrate the calibration of the design thresholds to ensure desirable frequentist operating characteristics in the selected scenario of interest. Similar to \hyperlink{R28}{Yin et al. (2012)}, when the null cases were considered, all entries with type I error rates were less or equal to 5\% in our study, for which the paired values of $(\lambda, \theta_{U})$ satisfied our requirements. Moreover, in alternative cases, we must find a paired value that provides a maximum power greater than 80\%.
Considering that we followed the aforementioned setting of parameters other than the pair of $(\lambda, \theta_{U})$, we conducted 10000 simulated clinical trials. 
Specifically, we evaluated the type I error and power values subject to the $H_{0}$ and $H_{1}$ by varying the design parameter $(\lambda, \theta_{U})$ when the underlying values of the (best response rate, best objective response rate) were equal to $(0.2, 0.2)$ with a correlation of  $0.5$ and $(0.7, 0.5)$ with a correlation of $0.5$ based on $H_{0}$ and $H_{1}$, respectively.
For each of the paired values of  $(\lambda, \theta_{U})$, we obtained the type I error and power, as listed in Table \hyperlink{T2}{2}.

\begin{center}
\lbrack Tables 2(a) and (b) about here.\rbrack
\end{center}

\noindent
In this case, the type I error rate was guaranteed to be less than $0.05$ except for the combinations $(\lambda, \theta_{U})=(0.65, 0.65)$, $(0.65, 0.70)$, $(0.65, 0.75)$ and $(0.70, 0.65)$ from Table \hyperlink{T2}{2}(a). One of these combinations can be selected to maximize the power using the boundary line of the staircase curve. Herein, Table \hyperlink{T2}{2}(b) indicates that it is appropriate to set $(\lambda, \theta_{U})=(0.65, 0.80)$, which is marked in bold, when designing a clinical trial.

\hypertarget{S4}{}
\section{Discussion}
This study proposed a new approach to incorporate the non-confirmed outcomes into the decision-making process. To avoid a substantial computational burden, one particularly important aspect was to obtain the quasi-posterior distribution from the closed form without using Monte Carlo methods. In addition, the operating characteristics of the proposed method were demonstrated based on a simple simulation study. Further, the calibration of the design parameters was addressed. Notably, the proposed design yielded desirable operating characteristics under certain conditions.  
Consequently, the proposed framework is expected to contribute to the effective identification of potent agents with sufficient activity for continued development. 
Further, our new approach may contribute to not only exploratory trials to confirm if the agent is worth continuing to the subsequent confirmatory trial, but also pivotal trials for the rare diseases, where the sample sizes required to perform clinical trials are usually limited.
\hyperlink{R26}{Ursino and Stallard (2021)} stated that the features of the Bayesian framework are extremely attractive and that the flexible Bayesian approach can be observed as the future gold standard in this field. In the development of rare diseases, the conduction of randomized trials and evaluation of the time-to-endpoints, such as progression-free survival or overall survival, is challenging.
Thus, a single-arm clinical trial with a response outcome as the primary endpoint can be planned as a pivotal trial. In practice, this has been implemented by \hyperlink{R8}{Fukano et al. (2020)}, \hyperlink{R16}{Nakamura et al. (2021)} and \hyperlink{R17}{O'Malley et al. (2022)}. Therefore, our new approach may also be valuable in this field.

In Section \hyperlink{S3}{3}, we confirmed that the proposed method based on the predictive probability was a powerful tool for making appropriate decisions.
However, in the context of the predictive probability, clinical trial designs using predictive probability for interim monitoring do not claim the magnitude of the treatment effect. Instead, the claim of treatment benefits is based on the Bayesian posterior probability. Similarly, it is important to consider the predictive probability as a decision-making framework as well as the Bayesian inference to evaluate the magnitude of the treatment effect in clinical trial designs. Although, we should at least report the posterior distribution of the difference between E and S, based on the confirmed outcome, the difference distribution based on the proposed approach may support the interpretation of the trial results. Based on this perspective, a Bayesian inference is provided in Appendix \hyperlink{A4}{D}.

Although the proposed method is considered to be a powerful framework for making appropriate decisions by optimizing the tuning parameters, a few limitations exist. First, we performed a simple simulation study to evaluate the PoC of the proposed approach. Because the findings of our simulation study were based on a specific simulation parameters and methods, generalization may be challenging. Further, the value of $w$ has a major impact on the results of the clinical trials; thus, clinicians and biostatisticians should collaborate intensively to obtain $w$ before conducting clinical trials. However, it may be difficult to set an optimal $w$ value in terms of regulatory aspects. Furthermore, the proposed approach assigned the same weights to responses without confirmation and to the cases with ongoing evaluations without initial response. Instead of assuming that these weights are common constant values, it may be beneficial to consider the augmentation of the design by using separate weights for non-confirmed outcomes w/o response.
Therefore, it is desirable to exploit a new design to estimate the values of $w$ from the obtained data in a more flexible manner and evaluate its behavior using a comprehensive simulation study.

\

\

\noindent
{\large\textit{Acknowledgements}}\\
The authors thank Takashi Asakawa, Yuki Nakagawa and Ryo Sawamoto of Chugai Pharmaceutical Co., Ltd. for their helpful comments and suggestions.

\

\noindent
{\large\textit{Author Contributions}}\\
Takuya Yoshimoto contributed to the study's conception and design. The analysis was performed by Takuya Yoshimoto, and all authors evaluated the results. The first draft of the manuscript was written by Takuya Yoshimoto, and all authors commented on previous versions of the manuscript. All the authors have read and approved the final version of this manuscript.

\

\noindent
{\large\textit{Funding}}\\
This research received no specific grants from funding agencies in the public, commercial, or non-for-profit sectors.

\

\noindent
{\large\textit{Conflict of Interest}}\\
Takuya Yoshimoto is an employee of Chugai Pharmaceutical Co., Ltd. The other authors declare no conflicts of interest.

\

\noindent
{\large\textit{Data Availability Statement}}\\
All source code can be requested from the corresponding author.

\

\noindent
{\large\textit{ORCID}}\\
Takuya Yoshimoto, \url{https://orcid.org/0000-0002-8747-1395}

\newpage
\hypertarget{A1}{}
\subsection*{Appendix A: Elicitation of Equation (2)}
Because the distributions of $p_{E}$ and $p_{S}$ are independent, the joint distribution is expressed as
$$
f \left(p_{E}, p_{S}|\alpha_{E}^{*}, \beta_{E}^{*}, \alpha_{S}, \beta_{S}\right)=beta(p_{E};\alpha_{E}^{*}, \beta_{E}^{*})beta(p_{S};\alpha_{S}, \beta_{S}).
$$
The experimental treatment considered sufficiently promising for a certain $p_{S} $ is calculated as
$$
\int_{p_{S}}^{1}f \left(p_{E}, p_{S}|\alpha_{E}^{*}, \beta_{E}^{*}, \alpha_{S}, \beta_{S}\right)dp_{E}=\frac{1}{B(\alpha_{S}, \beta_{S})}p_{S}^{\alpha_{S}-1}(1-p_{S})^{\beta_{S}-1}\left[1- Beta(p_{S}; \alpha_{E}^{*}, \beta_{E}^{*})\right],
$$
The probability that the experimental treatment is considered promising ${\rm Pr} \left(p_{E} > p_{S}|\alpha_{E}^{*}, \beta_{E}^{*}, \alpha_{S}, \beta_{S}\right)$ is derived to elicit the marginal distribution of $p_{S}$ as follows:
$$
{\rm Pr} \left(p_{E} > p_{S}|\alpha_{E}^{*}, \beta_{E}^{*}, \alpha_{S}, \beta_{S}\right)=\int_{0}^{1}\frac{1}{B(\alpha_{S}, \beta_{S})}p_{S}^{\alpha_{S}-1}(1-p_{S})^{\beta_{S}-1}\left[1- Beta(p_{S}; \alpha_{E}^{*}, \beta_{E}^{*})\right]dp_{S}.
$$

\


\hypertarget{A2}{}
\subsection*{Appendix B: Property of the bivariate Bernoulli distribution}

Let us assume that the probability distributions of the random variables $U_{1}$ and $U_{2}$ are bivariate Bernoulli. Let $p_{00}$ be the probability with $U_{1}=0$ and $U_{2}=0$, $p_{10}$ be the probability with $U_{1}=1$ and $U_{2}=0$, $p_{01}$ be the probability with $U_{1}=0$ and $U_{2}=1$, and $p_{11}$ be the probability with $U_{1}=1$ and $U_{2}=1$.
Subsequently, the probabilities $U_{1}=u_{1}$ and $U_{2}=u_{2}$ (denoted by $p_{u_{1}u_{2}}$) are expressed as follows:
$$
\displaystyle{\rm Pr}(U_{1}=u_{1}, U_{2}=u_{2})=p_{00}^{(1-u_{1})(1-u_{2})}p_{10}^{u_{1}(1-u_{2})}p_{01}^{(1-u_{1})u_{2}}p_{11}^{u_{1}u_{2}}.
$$
Consequently, the marginal probabilities of $U_{1}$ and $U_{2}$ are expressed as 
\begin{equation*}
\begin{split}
\displaystyle{\rm Pr}(U_{1}=u_{1})&=\sum_{u_{2}}p_{u_{1}u_{2}}=p_{00}^{1-u_{1}}p_{10}^{u_{1}}+p_{01}^{1-u_{1}}p_{11}^{u_{1}},\\[7pt]
\displaystyle{\rm Pr}(U_{2}=u_{2})&=\sum_{u_{1}}p_{u_{1}u_{2}}=p_{00}^{1-u_{2}}p_{01}^{u_{2}}+p_{10}^{1-u_{2}}p_{11}^{u_{2}}.
\end{split}
\end{equation*}
We now consider the $E(U_{1})$ and $E(U_{2})$,
\begin{equation*}
\begin{split}
\displaystyle{\rm E}(U_{1})&=\sum_{u_{1}}u_{1}{\rm Pr}(U_{1}=u_{1})=p_{10}+p_{11},\\[7pt]
\displaystyle{\rm E}(U_{2})&=\sum_{u_{2}}u_{2}{\rm Pr}(U_{2}=u_{2})=p_{01}+p_{11}.
\end{split}
\end{equation*}
For the variance of $U_{i}$, $i=1, 2$, let $E(U_{i})$ be denoted by $\pi_{i}$, 
\begin{equation*}
\begin{split}
\displaystyle{\rm Var}(U_{i})&=E(U_{i}^2)-E(U_{i})^2=\pi_{i}(1-\pi_{i}),\\[6pt]
\displaystyle{\rm E}(U_{1}U_{2})&=\sum_{u_{1}}\sum_{u_{2}}u_{1}u_{2}p_{u_{1}u_{2}}=p_{11}.
\end{split}
\end{equation*}
We can now derive the covariance of $(U_{1}, U_{2})$ as follows:
$$
\displaystyle{\rm Cov}(U_{1}, U_{2})={\rm E}(U_{1}U_{2})-{\rm E}(U_{1}){\rm E}(U_{2})=p_{11}-\pi_{1}\pi_{2}.
$$
Consequently, we obtain the following correlation, which is denoted by $\rho$,
$$
{\rm Corr}(U_{1},U_{2})=\frac{p_{11}-\pi_{1}\pi_{2}}{\sqrt{\pi_{1}(1-\pi_{1})\pi_{2}(1-\pi_{2})}},
$$
and the binary correlation matrix is expressed as
$$
\Sigma=
\begin{pmatrix}
{\sigma_{11}}&{\sigma_{12}} \\
{\sigma_{21}}&{\sigma_{22}}
\end{pmatrix},
$$
where $\sigma_{11}=\pi_{1}(1-\pi_{1})$, $\sigma_{22}=\pi_{2}(1-\pi_{2})$ and $\sigma_{12}=\sigma_{21}=\rho\sqrt{\sigma_{11}\sigma_{22}}$. For further details, refer to \hyperlink{R13}{Marshall and Olkin (1985)}. 

\


\hypertarget{A3}{}
\subsection*{Appendix C: Numerical algorithm to evaluate the operating characteristics of the proposed approach}

Herein, we explain the details of the simulation study discussed in Section \hyperlink{S3}{3} step by step. First, we should set suitable values for the parameters described in Section \hyperlink{S3}{3}, such as the $N_{min}$, $N_{max}$, number of cohort sizes, recruitment period and speed, tumor assessment period, timing of decision-making (e.g., after all the participants enrolled at the timing of interim monitoring have completed at least two tumor assessments), hyperparameter of E, parameter of S, and the number of simulations. Further, we set the probability thresholds $\lambda$ and $\theta_{U}$. In addition, scenarios should be assumed by setting the true values of the best response rate, the best objective response rate and correlation, which are denoted by $p_{BR}^{true}$, $p_{BOR}^{true}$ and $\rho^{true}$, respectively. 
Moreover, we should set an optimal weight for the non-confirmed outcome $w$ through the intensive discussions between clinicians and biostatisticians. Furthermore, we denote the current data of non-confirmed outcome as $x_{BR}$, and the current data of confirmed outcome as $x_{BOR}$.

Once these design parameters are fixed, a single trial with stopping rules can be simulated based on the proposed approach using the following steps:

\begin{itemize}
    \item[(a1)] Enroll the first $N_{min}$ participants into the trial by simulating the current data $x_{BR}$ and $x_{BOR}$ from a bivariate Bernoulli distribution with $N_{min}$, parameters $p_{BR}^{true}$, $p_{BOR}^{true}$ and $\rho^{true}$. Further, generate the time of enrollment, time to obtain the non-confirmed outcome and time to obtain the confirmed outcome from the uniform and normal distributions.

	\item[(a2)] From data created at the end of step (a1), calculate the decision-making timing. Specifically, we can identify the timing after all participants enrolled at the timing of interim monitoring at when they completed at least two tumor assessments.

	\item[(a3)] Re-map the binary outcome such as response or non-response depending on the timing of decision-making derived from step (a2). Namely, response within the decision time should be considered as a response. Otherwise, they should be considered as non-responders at that time.

     \item[(a4)] Compute $PP$ based on the current data of $N_{min}$ participants by using Equation (\hyperlink{E5}{5}). If $PP>\theta_{U}$ the algorithm stops. Whereas, if $PP \leq \theta_{U}$ the algorithm proceeds according to the following steps:

     \item[(a5)] Enroll the additional five participants (cohort size) into the trial in the same manner as that in step (a1). In addition, we should update the timing of decision-making, and re-map all binary outcomes for ($N_{min}$ + 5) participants based on the updated decision time.

     \item[(a6)] Compute $PP$ based on the current data of ($N_{min}$ + 5) participants by using Equation (\hyperlink{E5}{5}). If $PP>\theta_{U}$ the algorithm stops. Whereas, if $PP \leq \theta_{U}$ repeat the steps from (a5) to (a6).

     \item[(a7)]  If the maximum sample size $N_{max}$ has been reached, we should complete the trial.
\end{itemize}

We simulated 10000 trials by using the steps described above. Each simulated trial provided the value of the actually achieved sample size and study duration.
Based on the results of 10000 simulated trials, the performance metrics of PET, PRN, ASS and ASD were derived.

\


\hypertarget{A4}{}
\subsection*{Appendix D: Inference from the proposed method}
We here consider the differences between the two proportions. \hyperlink{R18}{Pham-Gia et al. (2017)} reported that the use of the Bayesian approach results in serious computational difficulties if a closed form expression for the density of the differences between two independent beta distributed random variables is not considered.
This expression has been discussed by \hyperlink{R19}{Pham-Gia et al. (1993)}; thus, we derive the two independent random variables of $p_{E}$ and $p_{S}$ at the time of the final analysis by using the existing theorem (see also, \hyperlink{R18}{Pham-Gia et al., 2017}).

First, \hyperlink{R19}{Pham-Gia et al. (1993)} introduced an expression by using the two-dimensional generalized hypergeometric function called Appell's hypergeometric function (denoted by $F_{1}$). $F_{1}$ can be expressed as a convergent integral:
$$
F_{1}(a,b_{1},b_{2};c;x_{1},x_{2})=\frac{\Gamma(c)}{\Gamma(a)\Gamma(c-a)}\int_{0}^{1}u^{a-1}(1-u)^{c-a-1}(1-ux_{1})^{-b_{1}}(1-ux_{2})^{-b_{2}}du,
$$
where $\Gamma(\cdot)$ denotes the gamma function. This converges when $c-a>0$, $a>0$. Then, the difference $p_{D}=p_{E}-p_{S}$ based on the proposed approach has a density defined in the range of $[-1, 1]$ as follows:
\[
f(p_{D}|\alpha_{E}^{*}, \beta_{E}^{*}, \alpha_{S}, \beta_{S})=
\renewcommand{\arraystretch}{1.3}
  \left\{
  \begin{array}{l@{}l@{}l}
         B(\alpha_{S}, &{}{}&\beta_{E}^{*})p_{D}^{\beta_{E}^{*}+\beta_{S}-1}(1-p_{D})^{\alpha_{S}+\beta_{E}^{*}-1}\\
		&{}{}&F_{1}(\beta_{E}^{*}, \alpha_{E}^{*}+\alpha_{S}+\beta_{E}^{*}+\beta_{S}-2, 1-\alpha_{E}^{*};\beta_{E}^{*}+\alpha_{S};1-p_{D}, 1-p_{D}^2)/A\\
		&\multicolumn{2}{r}{\left(0 <p_{D}\leq 1\right),}\\[7pt]
          \multicolumn{3}{l}{B(\alpha_{E}^{*} + \alpha_{S} - 1, \beta_{E}^{*}+\beta_{S}-1)/A}\\
          &\multicolumn{2}{r}{\left(p_{D}=0, \text{ if } \alpha_{E}^{*} + \alpha_{S} >1,~\beta_{E}^{*} + \beta_{S}>1\right),}\\[7pt]
         \multicolumn{3}{l}{B(\alpha_{E}^{*}, \beta_{S})(-p_{D})^{\beta_{E}^{*}+\beta_{S}-1}(1+p_{D})^{\alpha_{E}^{*}+\beta_{S}-1}}\\
         &{}{}&F_{1}(\beta_{S}, 1-\alpha_{S}, \alpha_{E}^{*}+\alpha_{S}+\beta_{E}^{*}+\beta_{S}-2; \alpha_{E}^{*}+\beta_{S};1-p_{D}^2, 1+p_{D})/A\\
         &\multicolumn{2}{r}{\left(-1\leq p_{D}<0\right),}
  \end{array}
  \right.
\]
where $A=B(\alpha_{E}^{*}, \beta_{E}^{*})B(\alpha_{S}, \beta_{S})$. Second, we describe the posterior mean and the $95\%$ credible intervals for $p_{D}$. The posterior mean ${\rm E}(p_{D}|\alpha_{E}^{*}, \beta_{E}^{*}, \alpha_{S}, \beta_{S})$ is 
$$
\displaystyle\int_{-1}^{1}p_{D}f(p_{D}|\alpha_{E}^{*}, \beta_{E}^{*}, \alpha_{S}, \beta_{S})dp_{D},
$$
and the $95\%$ credible interval $\left[l(p_{D}), u(p_{D})\right]$ is calculated as
\begin{equation*}
\begin{split}
\displaystyle\int_{-1}^{l(p_{D})}f(p_{D}|\alpha_{E}^{*}, \beta_{E}^{*}, \alpha_{S}, \beta_{S})dp_{D}&=0.025,\\[7pt]
\displaystyle\int_{u(p_{D})}^{1}f(p_{D}|\alpha_{E}^{*}, \beta_{E}^{*}, \alpha_{S}, \beta_{S})dp_{D}&=0.025.
\end{split}
\end{equation*}

Finally, we demonstrate the elicitation of the density for $p_{D}$ through a numerical example. We here assume that $\alpha_{E}^{*}=18.5$ and $\beta_{E}^{*}=22.5$ in the final analysis, with $\alpha_{S}=20$ and $\beta_{S}=80$. Figure \hyperlink{FA1}{A1} shows the distributions of $p_{D}$, $p_{E}$ and $p_{S}$, respectively. We can obtain a result with a posterior mean of $0.25$ and a $95\%$ credible interval of $[0.08, 0.42]$ for $p_{D}$.

\begin{center}
\lbrack Figure A1 about here.\rbrack
\end{center}
\noindent
We can easily report the inference on $p_{D}$ based on the confirmed outcome, as the posterior distribution can simply be updated using Bayes' theorem from the binary outcome. 
Notably, in terms of the calculation of the predictive probability, the elicitation of the distribution itself is not necessary.
Furthermore, it is sufficient to calculate the probability that the experimental treatment will be declared sufficiently promising, that is, Equation (\hyperlink{E4}{4}), to consider a well-known issue with the computational burden highlighted by \hyperlink{R22}{Saville et al. (2014)}.

\newpage
\hypertarget{T1}{}
\noindent
{\bf{Table 1:}}
Description of the scenarios of interest and the probability of early termination (PET), the proportion of rejection of the null hypothesis (PRN), the average sample size (ASS), and the average study duration (ASD), with 10000 simulated trials. BR and BOR indicate the best response rate and the best objective response rate, respectively. In addition, the term ``pts'' indicates the participants.

\

\renewcommand{\arraystretch}{1.5}
\begin{table}[htbp]
\centering
\begin{tabular} {lclcccc} 
\hline
Scenario & (BR, BOR)   & Methodology & ~~PET~~  & ~~PRN~~  &~~ASS~~  & ~~ASD~~  \\
             &                   &                    &               &               & (pts)   & (day) \\\hline  
1           &  $(0.7, 0.4) $ & Proposed approach                   & 0.4553 & 0.4574 & 29.1770  & 965.5258 \\
            &                    & Performance-oriented approach & 0.4579 & 0.4579 & 29.1360  & 1401.9155 \\
            &                    & Expedition-oriented approach     & 0.2020 & 0.2020 & 35.5215 & 1169.8997 \\ \cdashline{1-7}
2           &  $(0.7, 0.5) $ & Proposed approach                   & 0.7606 & 0.7703 & 22.0225  & 736.6134 \\
            &                    & Performance-oriented approach & 0.7968 & 0.7968 & 21.0435  & 954.5569 \\
            &                    & Expedition-oriented approach    & 0.5301 & 0.5301 & 28.5490  & 945.6213 \\ \cdashline{1-7}
3           &  $(0.7, 0.6) $ & Proposed approach                   & 0.9491 & 0.9587 & 17.1880  & 581.0368 \\
            &                    & Performance-oriented approach & 0.9656 & 0.9656 & 16.4690  & 700.0237 \\
            &                    & Expedition-oriented approach    & 0.8478 & 0.8478 & 21.2585  & 711.6206 \\ \cdashline{1-7}
4           &  $(0.4, 0.2) $ & Proposed approach                    & 0.0232 & 0.0233 &39.4295 & 1293.8817\\
            &                    & Performance-oriented approach & 0.0166 & 0.0166 & 39.5900  & 1983.1470\\
            &                    & Expedition-oriented approach    & 0.0043 & 0.0043 & 39.8950 & 1309.2621\\ \cdashline{1-7}
5           &  $(0.3, 0.2) $ & Proposed approach                   & 0.0233 & 0.0233 & 39.4270  & 1294.5907\\
            &                    & Performance-oriented approach & 0.0183 & 0.0183 & 39.5440 & 1982.4126\\
            &                    & Expedition-oriented approach    & 0.0048 & 0.0048 & 39.8840 & 1309.0559\\ \cdashline{1-7}
6           &  $(0.2, 0.2) $ & Proposed approach approach    & 0.0169  & 0.0169 & 39.5860 & 1299.6468\\
            &                    & Performance-oriented approach & 0.0201 & 0.0201 & 39.5025 & 1979.3308 \\
            &                    & Expedition-oriented approach    & 0.0038 & 0.0038 & 39.9075 & 1310.5946\\\hline
\end{tabular}
\end{table}

\newpage
\hypertarget{T2}{}
\noindent
{\bf{Table 2:}}
The type I error and power values based on the $H_{0}$ and $H_{1}$ hypotheses for the proposed approach by varying the design parameter $(\lambda, \theta_{U})$ when the (best response rate, best objective response rate) values are equal to $(0.2, 0.2)$ (with correlation $0.5$) and $(0.7, 0.5)$ (with correlation $0.5$) based on the $H_{0}$ and $H_{1}$ hypotheses (10000 simulated trials). 

\

\renewcommand{\arraystretch}{1.5}
\begin{table}[htbp]
\centering
\begin{tabular} {lcccccc} 
\hline
$\theta_{U}$ & \multicolumn{6}{c}{Null hypothesis with $\lambda$} \\\hline
        & 0.65  & 0.70 & 0.75 & 0.80 & 0.85 & 0.90 \\\hline
0.65  & 0.0747 & \multicolumn{1}{c|}{0.0643} & 0.0459 & 0.0342 & 0.0346 & 0.0159   \\ \cline{3-3}
0.70  & \multicolumn{1}{c|}{0.0661} & 0.0468 & 0.0416 & 0.0264 & 0.0211 & 0.0084  \\
0.75  & \multicolumn{1}{c|}{0.0671} & 0.0361 & 0.0330 & 0.0243 & 0.0114 & 0.0084  \\ \cline{2-2}
0.80  & \textbf{0.0403} & 0.0333 & 0.0176 & 0.0156 & 0.0085 & 0.0084 \\
0.85  & 0.0243 & 0.0235 & 0.0121 & 0.0074 & 0.0081 & 0.0032  \\
0.90  & 0.0176 & 0.0109 & 0.0101 & 0.0060 & 0.0036 & 0.0019   \\\hline
\end{tabular}
\captionsetup{labelformat=empty,labelsep=none}
\caption{(a) Type I error under $H_{0}$}
\end{table}

\

\begin{table}[htbp]
\centering
\begin{tabular} {lcccccc} 
\hline
$\theta_{U}$ & \multicolumn{6}{c}{Alternative hypothesis with $\lambda$} \\\hline
        & 0.65  & 0.70 & 0.75 & 0.80 & 0.85 & 0.90 \\\hline
0.65  & 0.9364 & 0.9204  & 0.8920 & 0.8500 & \multicolumn{1}{c|}{0.8186} & 0.7189  \\ \cline{6-6}
0.70  & 0.9277 & 0.8991 & 0.8854 & \multicolumn{1}{c|}{0.8283} & 0.7789 & 0.6630  \\
0.75  & 0.9233 & 0.8842 & 0.8571 & \multicolumn{1}{c|}{0.8135} & 0.7393 & 0.6436  \\ \cline{5-5}
0.80  & \textbf{0.9008} & 0.8688 & \multicolumn{1}{c|}{0.8134} & 0.7794 & 0.6979 & 0.6222  \\ \cline{4-4}
0.85  & 0.8723 & \multicolumn{1}{c|}{0.8467} & 0.7891 & 0.7201 & 0.6500 & 0.5438  \\
0.90  & 0.8514 & \multicolumn{1}{c|}{0.8032} & 0.7600 & 0.6804 & 0.5965 & 0.4612  \\\hline
\end{tabular}
\captionsetup{labelformat=empty,labelsep=none}
\caption{(b) Statistical power under $H_{1}$}
\end{table}

\newpage
\hypertarget{FA1}{}
\noindent
{\bf{Figure A1:}}
Density functions of $p_{D}$, $p_{E}$ and $p_{S}$ are assumed to be obtained $\alpha_{E}^{*}=18.5$ and $\beta_{E}^{*}=22.5$ in the final analysis, with $\alpha_{S}=20$ and $\beta_{S}=80$.
The symbols E and S represent the experimental and standard treatments, respectively, and D represents the difference between E and S.

\

\begin{figure}[h]
\centering
\includegraphics[scale=0.5]{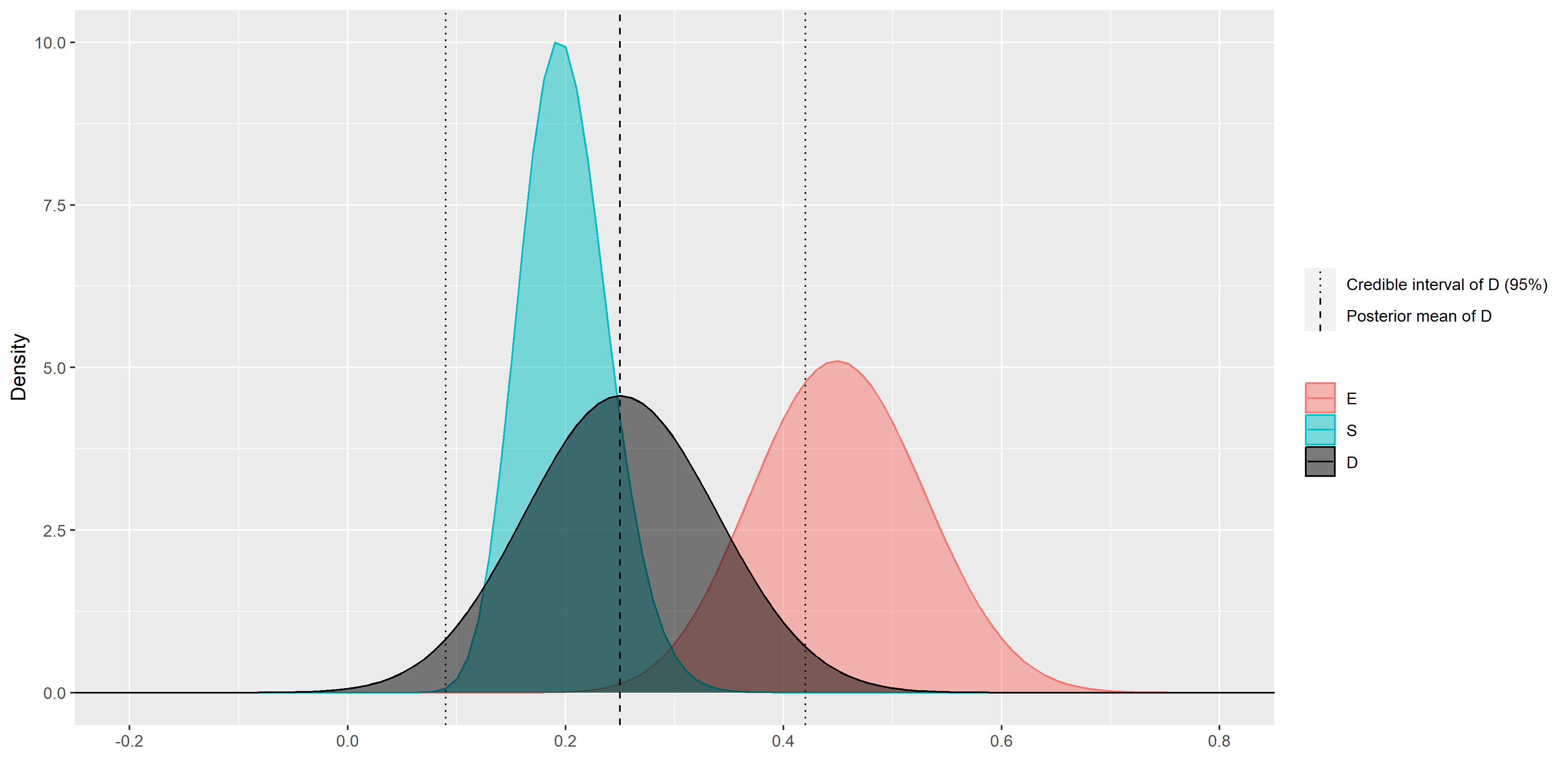}
\captionsetup{labelformat=empty,labelsep=none}
\end{figure}

\end{document}